\documentclass[aip,graphicx,groupedaddress]{revtex4-1}

\usepackage{graphicx}
\usepackage{amsmath, amssymb, nccmath, appendix, comment}
\usepackage{hyperref, physics}
\usepackage{dcolumn}
\usepackage{bm}

\usepackage[utf8]{inputenc}
\usepackage[T1]{fontenc}
\usepackage{mathptmx}
\usepackage{etoolbox}

\newcommand{\cA}{\mathcal{A}}

\draft

\begin{document}

\title{When action is not least for systems with action-dependent Lagrangians} 

\author{Joseph Ryan}
\email[]{jwryan@mail.smu.edu}
\affiliation{Department of Physics, Southern Methodist University, Dallas, TX 75205, USA}

\date{\today}

\begin{abstract}
The dynamics of some non-conservative and dissipative systems can be derived by calculating the first variation of an action-dependent action, according to the variational principle of Herglotz. This is directly analogous to the variational principle of Hamilton commonly used to derive the dynamics of conservative systems. In a similar fashion, just as the second variation of a conservative system's action can be used to infer whether that system's possible trajectories are dynamically stable, so too can the second variation of the action-dependent action be used to infer whether the possible trajectories of non-conservative and dissipative systems are dynamically stable. In this paper I show, generalizing earlier analyses of the second variation of the action for conservative systems, how to calculate the second variation of the action-dependent action and how to apply it to two physically important systems: a time-independent harmonic oscillator and a time-dependent harmonic oscillator.
\end{abstract}


\maketitle

\section{Introduction}

The principle of stationary action is a powerful unifying principle for physics. The dynamics of any conservative system, whose configuration at a given time $t$ is described by the generalized coordinates $\vec{x}(t) = \left(x_1(t), x_2(t), ... , x_n(t)\right)^{\rm T}$ and the generalized velocity $\vec{v}(t) = \frac{d\vec{x}(t)}{dt}$, can be derived by demanding that the functional
\begin{equation}
    \label{eq:action_def}
    S = \int_{t_i}^{t_f} L\left(\vec{x}(t), \vec{v}(t)\right)dt,
\end{equation}
be stationary with respect to variations of the form
\begin{equation}
\label{eq:x_var}
    \vec{x}\left(t, \epsilon\right) = \vec{x}\left(t, 0\right) + \epsilon \vec{y}\left(t\right),
\end{equation}
where $y(t)$ satisfies $y\left(t_i\right) = y\left(t_f\right) = 0$ but is otherwise arbitrary, and $\epsilon$ is an arbitrary infinitesimal parameter.\cite{Goldstein_Poole_Safko_2002, Fowles_Cassiday_2007} By ``stationary'', I mean that
\begin{equation}
\label{eq:dS/de=0}
    \frac{dS}{d\epsilon}\bigg|_{\epsilon \rightarrow 0} = 0.
\end{equation}
The system will follow the trajectories that make the action stationary as defined by Eq. (\ref{eq:dS/de=0}). For this condition to be satisfied, the Lagrangian $L\left(\vec{x}(t), \vec{v}(t)\right)$ must obey the Euler-Lagrange equations
\begin{equation}
\label{eq:Conservative_E-L_eqs}
    \frac{d}{dt}\frac{\partial L}{\partial \dot{x}_i} = \frac{\partial L}{\partial x_i},
\end{equation}
where $x_i$ refers to the $i^{\rm th}$ component of the vector $\vec{x}(t)$ and $\dot{x}_i := v_i$. The trajectories that make the action stationary must therefore be the ones that satisfy Eq. (\ref{eq:Conservative_E-L_eqs}). In non-relativistic mechanics, the Lagrangian takes the form $L = K - U$, where $K$ is the kinetic energy of the system and $U$ is the potential energy of the system. Finally, it is important to note that the Euler-Lagrange equations are valid for any time $t_i \leq t \leq t_f$, and not just at the endpoints $t_i$ and $t_f$ (this will be important in Sec. \ref{subsec:Herglotz_second_var}).

This variational, energy-based approach to mechanics can be a very powerful alternative to a force-based approach, and it is more easily generalized to the quantum domain. \cite{Fowles_Cassiday_2007} As formulated above, however, it cannot be applied to non-conservative systems. There has been much discussion of this point in the literature (see, for example, Refs. \citenum{Fitzgerald_et_al_EJMBP_2021, Riewe_PRE_1996, Galley_PRL_2013, Lemos_AJP_1981, Musielak_JPhA_2008, Denman_AJP_1966, Wang_Wang_arXiv_2012, Lin_Wang_2013, Udwadia_Leitmann_Cho_JAM_2011, Kobussen_Austriaca_1979, Bahar_Kwatny_AJP_1981, Langley_JMP_2021, Leitmann_Remarks_1963}) and many techniques have been proposed for handling non-conservative systems within a variational framework. One technique is to use a Lagrangian that depends explicitly on time such that $L = L\left(t, \vec{x}, \vec{v}\right)$, to incorporate the effects of, for example, time-dependent external fields or time-dependent constraints.\cite{Gray_Karl_Novikov_RPP_2004} Dissipation that is linear in the velocity can be accounted for by using a Lagrangian of the form $L = e^{kt}\left(K - U\right)$, where $k$ is a constant, although there is a long-running debate in the literature on whether this produces physically meaningful energy and momentum functions (see, for example, Refs. \citenum{Fitzgerald_et_al_EJMBP_2021, Lin_Wang_2013, Lazo_Krumreich_JMP_2014, Musielak_JPhA_2008, Riewe_PRE_1996, Tartaglia_EJP_1983, Kobe_Reali_Sieniutycz_AJP_1986, Lemos_AJP_1981, Ray_AJP_1979, Greenberger_JMP_1979, Edwards_AJP_1979} and references therein). Another way to handle non-conservative and dissipative systems, which does produce physically meaningful energy and momentum functions, is to use a Lagrangian that is an explicit function of the action $S$ (it may also be an explicit function of time).\cite{Lazo_et_al_JMP_2018} This approach was developed by Gustav Herglotz,\cite{Herglotz_Contact_Lectures} and it has recently been explored by several authors (see, for example, Refs. \citenum{Georgieva_Guenther_Bodurov_JMP_2003, Lazo_et_al_PRD_2017, Lazo_et_al_JMP_2018, Paiva_Lazo_Zanchin_arXiv_2021, Georgieva_Guenther_2005, Zhang_Tian_PLA_2019, Tian_Zhang_Adiabatic_2020, Sloan_Cosmology_2021, Donchev_JMP_2014}). A Lagrangian that depends explicity on $S$ can be used to derive equations of motion that describe certain types of non-conservative and dissipative systems, and it contains the standard conservative Lagrangian as a special case. Some investigators have produced generalized versions of Noether's theorems stemming from the symmetries of the Herglotz Lagrangian,\cite{Georgieva_Guenther_Bodurov_JMP_2003, Georgieva_Guenther_2005, Zhang_Tian_PLA_2019} and have used these to discover conservation laws for non-conservative systems \cite{Donchev_JMP_2014} (see also Ref. \citenum{Tian_Zhang_Adiabatic_2020} for an analysis of adiabatic invariants for non-conservative systems, and Ref. \citenum{Gaset_et_al_IJGMMP_2020} for a geometric approach to the Noether symmetry analysis). Others have broadened the applicability of the Herglotz Lagrangian to field theory, in both its non-covariant \cite{Georgieva_Guenther_Bodurov_JMP_2003} and its covariant\cite{Lazo_et_al_PRD_2017, Lazo_et_al_JMP_2018} forms, leading to alternative theories of gravity and cosmology. \cite{Lazo_et_al_PRD_2017, Paiva_Lazo_Zanchin_arXiv_2021, Sloan_Cosmology_2021, Carames_et_al_PRD_2018} The Herglotz action principle is also closely linked to the study of contact geometry, a generalization of the symplectic geometry associated with conservative Hamiltonian systems, which can be used to describe non-conservative interactions (see, for example, Refs. \citenum{Bravetti_Entropy_2017, Bravetti_Cruz_Tapias_Ann_Phys_2017, Ciaglia_Cruz_Marmo_Ann_Phys_2018, Gaset_et_al_Ann_Phys_2020, Gaset_et_al_RMP_2021, Bravetti_Jackman_Sloan_arXiv_2022, de_Leon_et_al_Monatshefte_2022})

The stationary action principle, whether it is applied to conservative or to non-conservative systems, is sometimes called the ``least action principle'', but this can be misleading in some situations. It is not always the case that the action of a conservative system is minimized, because Eq. (\ref{eq:dS/de=0}) can pick out a saddle point as well as a minimum (and, as shown in Ref. \citenum{Gray_Taylor_2007}, the action never attains a true maximum). To answer the question of whether the system's action has a true minimum or a saddle point requires the calculation of the \textit{second} variation of the system's action along a stationary path, and it turns out to be both interesting and useful to do this. In classical mechanics, the second variation of a system's action has a close connection to its dynamical stability, which I will explain in more detail below.\cite{Papstavridis_JSV_1982, Papastavridis_1983, Papastavridis_JAM_1985} Additionally, in quantum mechanics, the phase of the semiclassical propagator $K(t_i, t_f)$ is determined by the sign of the second variation of the classical action.\footnote{See Ref. \citenum{Levit_Smilansky_AnnPhys_1977}. Technically the phase is equal to $-\nu \pi/2$, where $\nu$ is the number of negative eigenvalues of the operator $\Lambda_{ij}$, which I define in Eq. (\ref{eq:Lambda_def}).} Analyses of the second variation of the action for conservative, classical systems have been carried out in Refs. \citenum{Gray_Taylor_2007, Gray_Poisson_AJP_2011, Hussein_et_al_1980, Papastavridis_1983}, and some studies of the second variation of the action for non-conservative systems exist (see, for example, Refs. \citenum{Papstavridis_JSV_1982, Papastavridis_JAM_1985, Papastavridis_Chen_JSV_1986, Gray_Taylor_2007}) but as far as I know the dynamical stability properties associated with the second variation of the Herglotz action have not been investigated. An explicit calculation of the second variation of the Herglotz action, similar to what I derive in Sec. \ref{subsec:Herglotz_second_var}, can be found in Ref. \citenum{Cheng_Hong_arXiv_2021}. As far as I can tell, however, the authors of that paper appear to have been concerned only with proving that the solution of the Herglotz variational problem minimizes the Herglotz action, and not with investigating the sign change of the second variation. Additionally, these authors do not appear to make the assumption, as I do, that the second variation of the Herglotz action vanishes at the initial time $t_i$. My goal, therefore, is to generalize the earlier analyses of the second variation of the action for conservative systems, focusing mainly on Refs. \citenum{Gray_Taylor_2007, Hussein_et_al_1980}. I will do this by:

1.) showing that the sign of the second variation of the Herglotz action can be related to the dynamical stability of non-conservative systems, following Ref. \citenum{Gray_Taylor_2007}, and

2.) showing that, with a few changes to the relevant equations, the sign of the second variation of the action can be calculated by way of a linear eigenvalue equation, in the same manner as for a conservative system (following Ref. \citenum{Hussein_et_al_1980}).

In Sec. \ref{sec:Herglotz_VP}, I summarize the Herglotz variational principle, starting with the first variation in Sec. \ref{subsec:Herglotz_first_var}, moving on to the second variation in Sec. \ref{subsec:Herglotz_second_var}. I then derive the connection between the sign of the second variation of the Herglotz action and dynamical stability in Sec. \ref{sec:3}. Sec. \ref{sec:eig_analysis} describes a method for determining the sign of the second variation of the Herglotz action, followed by two simple examples in Secs. \ref{subsec:SHO_a} and \ref{subsec:SHO_g}. Throughout this paper I use the Einstein summation convention for repeated indices. For readers who may wish to skip the lengthy derivations, the following list provides a summary of the key material:

\begin{itemize}

    \item{Euler-Lagrange equation for a system described by the Herglotz action, Eq. (\ref{eq:EoM}).}

    \item{Second variation of the Herglotz action, Eq. (\ref{eq:f'=ver2}), followed by comparison to the second variation of the standard conservative action in Eq. (\ref{eq:f'=consv}).}

    \item{Linear eigenvalue equation for the action eigenvalue spectrum of a one-dimensional system with time-dependent dissipation, Eq. (\ref{eq:short_eigenvalue_equation}).}

    \item{Action eigenvalue spectrum of a simple harmonic oscillator with time-independent dissipation, Eq. (\ref{eq:SHO_action_eigv_spec}).}

    \item{Numerical estimate of the lowest order action eigenvalue of a simple harmonic oscillator with time-dependent dissipation and time-dependent frequency, as a function of final time $t_f$, Fig. \ref{fig:Ground_eig_var_est_python}. Validation of this result, using a numerical solution of the equations of motion for the harmonic oscillator with time-dependent dissipation and time-dependent frequency, Fig. \ref{fig:TDHO_x}.}

\end{itemize} 

\section{Herglotz variational principle}
\label{sec:Herglotz_VP}
\subsection{First variation}
\label{subsec:Herglotz_first_var}
To obtain the equation(s) of motion, it is necessary to calculate the functional derivative, not of an integral equation, but of a differential equation
\begin{equation}
\label{eq:dSdt}
    \frac{dS}{dt} = L\left(t, x_i, \dot{x}_i, S\right),
\end{equation}
with the initial condition on the action being $S(t_i = 0) =: S_{t_i}$.\cite{Herglotz_Contact_Lectures} The reader should note here that the action $S$ defined in Eq. (\ref{eq:dSdt}) depends on the time $t$, unlike the action $S$ defined in Eq. \ref{eq:action_def}, which does not (assuming $t_f$ and $t_i$ are not allowed to vary). In the rest of this paper, $t_i = 0$. We want $S$ to be stationary with respect to variations of the form shown in Eq. (\ref{eq:x_var}), and by ``stationary'' I mean that
\begin{equation}
    \frac{dS}{d\epsilon}\bigg|_{\{\epsilon \rightarrow 0, \hspace{1mm} t = t_f\}} = 0,
\end{equation}
which is similar to, but not identical with, the condition that conservative systems must satisfy. In contrast to the standard conservative action, the first variation of the action vanishes only at the final time $t_f$.\cite{Herglotz_Contact_Lectures} Taking the functional derivative of Eq. (\ref{eq:dSdt}), we find
\begin{equation}
\label{eq:ddSdtdep}
\begin{aligned}
    \frac{d}{d\epsilon}\left[\frac{dS}{dt}\right]\bigg|_{\epsilon \rightarrow 0} & = \frac{d}{dt}\frac{dS}{d\epsilon}\bigg|_{\epsilon \rightarrow 0}\\
    & = \frac{\partial L}{\partial x_i}y_i + \frac{\partial L}{\partial \dot{x}_i}\dot{y}_i + \frac{\partial L}{\partial S}\frac{dS}{d\epsilon}\bigg|_{\epsilon \rightarrow 0},\\
\end{aligned}
\end{equation}
where the partial derivatives of $L$ are evaluated in the limit $\epsilon \rightarrow 0$ (that is, on the trajectory that solves the equations of motion). In the first line of Eq. (\ref{eq:ddSdtdep}), the operators $\frac{d}{dt}\frac{d}{d\epsilon}$ commute because $\epsilon$ is independent of time. If we define
\begin{equation}
\label{eq:f_def}
    f := \frac{dS}{d\epsilon}\bigg|_{\epsilon \rightarrow 0},
\end{equation}
and
\begin{equation}
\label{eq:B_def}
    B := \frac{\partial L}{\partial S},
\end{equation}
then Eq. (\ref{eq:ddSdtdep}) becomes
\begin{equation}
\label{eq:ddSdtdep_2}
    \frac{df}{dt} = \frac{\partial L}{\partial x_i}y_i + \frac{\partial L}{\partial \dot{x}_i}\dot{y}_i + Bf.
\end{equation}
This equation has the form $\frac{df}{dt} = \left(...\right) + Bf$, suggesting that the solution should be something like
\begin{equation}
\label{eq:f=Aexp+const}
    f = Ae^{\int B\left(\tau\right) d\tau} + {\rm const},
\end{equation}
where $A$ is an undetermined function of time. The derivative of $f$ is
\begin{equation}
\label{eq:dfdt=dAdt+Bf}
    \frac{df}{dt} = \frac{dA}{dt}e^{\int B\left(\tau\right) d\tau} + Bf,
\end{equation}
and by comparing this equation to Eq. (\ref{eq:ddSdtdep_2}), we can see that Eq. (\ref{eq:f=Aexp+const}) solves Eq. (\ref{eq:ddSdtdep_2}) if
\begin{equation}
\label{eq:dAdt_def}
    \frac{dA}{dt} := \left(\frac{\partial L}{\partial x_i}y_i + \frac{\partial L}{\partial \dot{x}_i}\dot{y}_i\right)e^{-\int B(\tau)d\tau}
\end{equation}
More formally, the solution of Eq. (\ref{eq:ddSdtdep_2}) is given by
\begin{equation}
\label{eq:f=Aexp+f0}
    f(t_f) = Ae^{\int_0^{t_f} B(\tau)d\tau} + f_i,
\end{equation}
where $f_i := \frac{d}{d\epsilon}S_{t_i}\big|_{\epsilon \rightarrow 0}$.

Because $S_{t_i}$ is independent of $x$, it follows that $\frac{d^n}{d\epsilon^n}S_{t_i}\big|_{\epsilon \rightarrow 0} = 0$, where $n$ is an integer, so $f_i = 0$.\cite{Vermeeren_Bravetti_Sero_JPA_2019} At time $t_f$, we demand $\frac{dS}{d\epsilon}\big|_{\epsilon \rightarrow 0} = 0$, so $f(t_f)$ vanishes and Eq. \ref{eq:f=Aexp+f0} becomes
\begin{equation}
\label{eq:0=}
    0 = \int_0^{t_f} \left(\frac{\partial L}{\partial x_i}y_i + \frac{\partial L}{\partial \dot{x}_i}\dot{y}_i\right)e^{-\int_0^{\tau_2} B(\tau_1)d\tau_1}d\tau_2.
\end{equation}
Integrating the second term in parentheses in Eq. (\ref{eq:0=}) by parts, and using the fact that $y_i(0) = y_i(t_f) = 0$ to discard the boundary terms, we find
\begin{equation}
\begin{aligned}
    0 = & \int^{t_f}_0 \frac{\partial L}{\partial x_i}y_i e^{-\int_0^{\tau_2} B(\tau_1)d\tau_1}d\tau_2\\
    & - \int^{t_f}_0 \left[\frac{d}{dt}\frac{\partial L}{\partial \dot{x}_i}y_i e^{-\int_0^{\tau_2} B(\tau_1)d\tau_1} - B\frac{\partial L}{\partial \dot{x}_i}y_i e^{-\int_0^{\tau_2} B(\tau_1)d\tau_1}\right]d\tau_2,\\
\end{aligned}
\end{equation}
which requires
\begin{equation}
\label{eq:EoM}
     \frac{d}{dt}\frac{\partial L}{\partial \dot{x}_i} = \frac{\partial L}{\partial x_i} + \frac{\partial L}{\partial S}\frac{\partial L}{\partial \dot{x}_i}.
\end{equation}
Eq. (\ref{eq:EoM}) is the generalized Euler-Lagrange equation for a system with non-conservative dynamics, described by an action-dependent Lagrangian. Applications of this generalized Euler-Lagrange equation can be found in, for example, Refs. \citenum{Lazo_et_al_JMP_2018, Lazo_et_al_PRD_2017, Paiva_Lazo_Zanchin_arXiv_2021, Sloan_Cosmology_2021, Georgieva_Guenther_2005, Georgieva_Guenther_Bodurov_JMP_2003, Gaset_Marin-Salvador_arXiv_2021, Carames_et_al_PRD_2018}.

\begin{figure*}
\centering
    \resizebox{\columnwidth}{!}{%
    \includegraphics[scale=1]{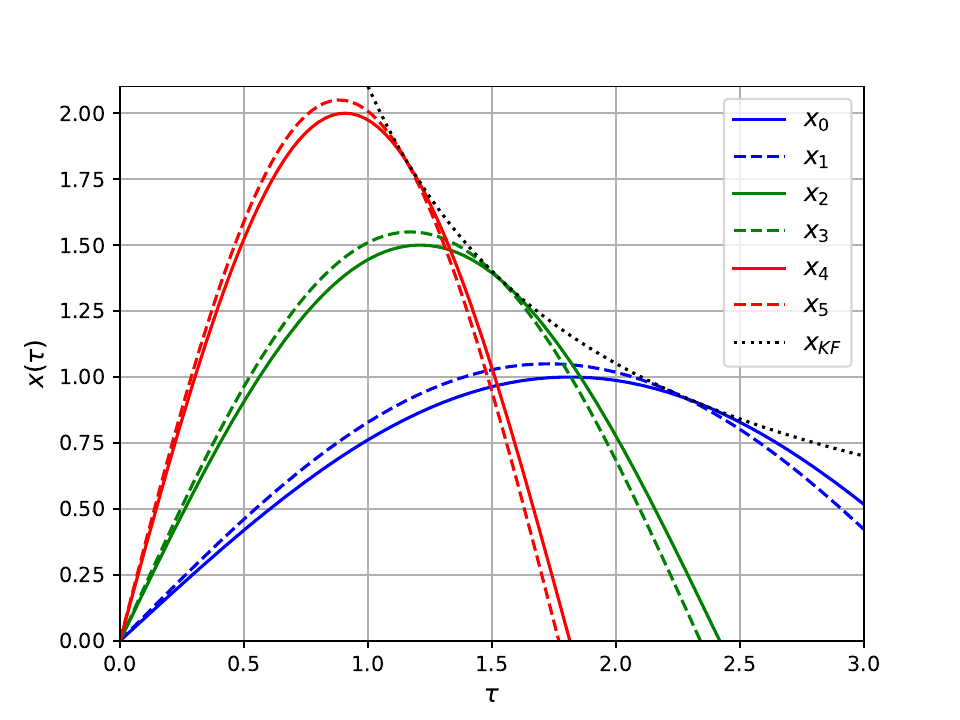}%
    }
\caption{Solid and dashed curves of the same color have slightly different amplitudes (see Table \ref{tab:QO_Initial_conditions}), and their points of intersection (kinetic foci) lie on the dotted black curve (the caustic). Displacements recorded on the vertical axis are measured in arbitrary units, and the times recorded on the horizontal axis have been scaled by a factor of $\sqrt{\frac{C}{m}}$.}
\label{fig:QO_KF}
\end{figure*}

For conservative systems, the first variation of the action is typically used to derive the equation(s) of motion, and the second variation of the action can be used to infer whether the system is dynamically stable, as I mentioned above.\cite{Gutzwiller_1990, Gray_Taylor_2007, Papastavridis_1983} To see how this works, consider a trajectory $\vec{x}_0(t)$ that satisfies a given system's equation(s) of motion. As shown in Ref. \citenum{Gray_Taylor_2007}, the second variation of the action for a conservative system is minimized, for all variations $\{\epsilon \vec{y}(t)\}$ away from the stationary path, as long as $t_f < t_{\rm KF1}$, where $t_f$ is the final time and ``KF1" stands for ``first kinetic focus". As defined in Refs. \citenum{Gray_Taylor_2007} and \citenum{Gray_Poisson_AJP_2011}, a kinetic focus (also known as a ``conjugate point''\cite{Gray_Poisson_AJP_2011}) is a point along $\vec{x}_0(t)$ at which a nearby trajectory coalesces with it in the limit that the difference of the initial velocities of both trajectories vanishes. More precisely, a kinetic focus is a point in time on $\vec{x}_0(t)$ that satisfies
\begin{equation}
\label{eq:kinetic_focus}
    {\rm det}\left[\frac{\partial \vec{x}_0(t)}{\partial \vec{v}_0}\right] = 0,
\end{equation}
where $\vec{v}_0$ is the initial velocity of the particle following $\vec{x}_0(t)$ (note that, for one-dimensional trajectories, Eq. (\ref{eq:kinetic_focus}) reduces to $\frac{\partial x_0(t)}{\partial v_0} = 0$). From these equations, we can see that kinetic foci play an important role in characterizing the dynamical stability of a given system. In a dynamically unstable system, two nearby trajectories with similar initial velocities will move farther apart as time goes on. The possible trajectories spread, or ``fan out'' \cite{Gutzwiller_1990, Gray_Taylor_2007} in a dynamically unstable system, whereas the possible trajectories of a system that is dynamically stable will converge, in the sense described above.\cite{Gutzwiller_1990, Gray_Taylor_2007} When $\vec{x}_0(t)$ reaches its first kinetic focus, the second variation of the system's action will be minimized for all but one of the variations away from $\vec{x}_0(t)$. This special variation (call it $\epsilon \vec{y}_{*}(t)$) will be another true trajectory of the system, because it satisfies the system's equation(s) of motion.\cite{Gray_Taylor_2007, Gray_Poisson_AJP_2011} Further, when $t_f > t_{\rm KF1}$, the second variation of the system's action will be maximized with respect to $\epsilon \vec{y}_{*}(t)$, and minimized with respect to all other variations (until the system reaches another kinetic focus; for more details see Refs. \citenum{Gray_Taylor_2007, Gray_Poisson_AJP_2011}). In general, then, the action of a conservative system is minimized if $t_f < t_{\rm KF1}$, and is a saddle point if $t_f > t_{\rm KF1}$.\cite{Gray_Taylor_2007, Papastavridis_1983}

Roughly speaking, then, a kinetic focus is a kind of ``spacetime focal point" for trajectories having similar initial velocities. A visual example of kinetic foci is provided by Fig. \ref{fig:QO_KF}. In this figure, following the example of Fig. 4 in Ref. \citenum{Gray_Taylor_2007}, I show several possible trajectories of a quartic oscillator (that is, an oscillator with $U = \frac{1}{4}Cx^4$). As in Refs. \citenum{Gray_Taylor_2007} and \citenum{Gray_Karl_Novikov_AJP_1996}, I use approximate solutions of the form
\begin{equation}
    x \approx A {\rm sin}\left(\omega_0 t\right),
\end{equation}
where
\begin{equation}
    \omega_0 \approx \sqrt{\frac{3C}{4m}}A.
\end{equation}
Redefining the time variable to have the form
\begin{equation}
    \tau := \sqrt{\frac{C}{m}}t,
\end{equation}
gives
\begin{equation}
    x \approx A{\rm sin}\left(\sqrt{\frac{3}{4}}A\tau\right).
\end{equation}
The trajectories shown in Fig. \ref{fig:QO_KF} differ in their amplitudes according to Table \ref{tab:QO_Initial_conditions}. The important thing to note here is that solid and dashed curves of the same color have amplitudes that are only slightly different. Because $v_0 = \omega_0 A$, a slight change in the amplitude of a given trajectory corresponds to a slight change in that trajectory's initial velocity. From Fig. \ref{fig:QO_KF}, we can see that trajectories having similar initial velocities eventually intersect; these intersection events are called kinetic foci.\footnote{The black dotted curve in the figure is known as the \textit{caustic}, and it is an envelope that divides spacetime into regions that are accessible from the given initial conditions, and regions that aren't (see Ref. \citenum{Gray_Taylor_2007}). Generally speaking, the caustic is composed of the kinetic foci associated with the family of trajectories that are dynamically allowed within a given potential (i.e. the family of solutions to the system's equation or equations of motion). I will not be concerned with caustics in the rest of this paper, because the examples I consider in Secs. \ref{subsec:SHO_a} and \ref{subsec:SHO_g}, namely harmonic oscillators with time-independent and time-dependent damping and frequency, do not exhibit caustics. Harmonic oscillators are special in that trajectories with arbitrarily large differences in their initial velocities reach the same kinetic focus, so for these systems the caustic curve collapses to a point \cite{Gray_Taylor_2007}.} The reader should note that I have only shown the first kinetic foci associated with the trajectories in Fig. \ref{fig:QO_KF}, and it is at these points that the second variation of the action, evaluated on a particular trajectory, vanishes for the first time. 
\begin{table*}
    \centering
    \begin{tabular}{c|c}
        \hline
         Quartic oscillator trajectories & $A$ (arb. units)\\
         \hline
         $x_0$ & 1\\
         $x_1$ & 1.05\\
         $x_2$ & 1.5\\
         $x_3$ & 1.55\\
         $x_4$ & 2\\
         $x_5$ & 2.05\\
         \hline
    \end{tabular}
    \caption{Trajectories used to generate Fig. \ref{fig:QO_KF}. The left column lists the trajectories with their corresponding label subscripts, and the right column lists the initial values of their amplitudes in arbitrary units.}
    \label{tab:QO_Initial_conditions}
\end{table*}
I will derive a general expression for the second variation of the Herglotz action in Sec. \ref{subsec:Herglotz_second_var}, which I will use in Sec. \ref{sec:3} to generalize the analysis of Ref. \citenum{Gray_Taylor_2007}.

\subsection{Second variation}
\label{subsec:Herglotz_second_var}
To calculate the second functional derivative of the Herglotz action, we take the functional derivative of Eq. (\ref{eq:ddSdtdep}):
\begin{equation}
\begin{aligned}
\label{eq:dddSdepdepdt}
    \frac{d}{d\epsilon}\left[\frac{d}{d\epsilon}\frac{dS}{dt}\right]\bigg|_{\epsilon \rightarrow 0} & = \frac{d}{d\epsilon}\left[\frac{d}{dt}\frac{dS}{d\epsilon}\right]\bigg|_{\epsilon \rightarrow 0} \\
    & = \frac{d}{d\epsilon}\left[\frac{df}{dt}\right]\bigg|_{\epsilon \rightarrow 0}\\
    & = \left[\frac{d\tilde{A}}{d\epsilon}\right]\bigg|_{\epsilon \rightarrow 0} + \left[\frac{dB}{d\epsilon}\right]\bigg|_{\epsilon \rightarrow 0}f + B\left[\frac{df}{d\epsilon}\right]\bigg|_{\epsilon \rightarrow 0},\\
\end{aligned}
\end{equation}
where $B$ and $f$ were defined in Eqs. (\ref{eq:B_def}) and (\ref{eq:f_def}), and
\begin{equation}
    \tilde{A} := \frac{\partial L}{\partial x_i}y_i + \frac{\partial L}{\partial \dot{x}_i}\dot{y}_i.
\end{equation}
Let $\frac{d(...)}{d\epsilon}\Big|_{\epsilon \rightarrow 0} := (...)'$. Then Eq. (\ref{eq:dddSdepdepdt}) becomes
\begin{equation}
\label{eq:df'/dt}
    \begin{aligned}
    \frac{d}{d\epsilon}\left[\frac{df}{dt}\right]\bigg|_{\epsilon \rightarrow 0} & = \frac{df'}{dt}\\
    & = \tilde{A}' + B'f + Bf'.\\
    \end{aligned}
\end{equation}
Eq. (\ref{eq:df'/dt}) is clearly similar to Eq. (\ref{eq:ddSdtdep_2}), and it can be solved in a similar fashion. Define
\begin{equation}
\label{eq:dG/dt_def}
\frac{dG}{dt} := \left(\tilde{A}' + B'f\right)e^{-\int B d\tau}
\end{equation}
so that
\begin{equation}
\label{eq:f'=}
    \begin{aligned}
    f' & = Ge^{\int B d\tau}\\
    & = \left[\int\left(\tilde{A}' + B'f\right)e^{-\int B d\tau_1} d\tau_2\right]e^{\int B d\tau_3}\\
    & = \Bigg[\int\Bigg(\frac{\partial^2 L}{\partial x_i \partial x_j}y_i y_j + 2\frac{\partial^2 L}{\partial x_i\partial\dot{x}_j}y_i\dot{y}_j  + \frac{\partial^2 L}{\partial \dot{x}_i \dot{x}_j}\dot{y}_i \dot{y}_j + 2\frac{\partial^2 L}{\partial S\partial \dot{x}_i}f\dot{y}_i\\
    & + 2\frac{\partial^2 L}{\partial x_i\partial S}fy_i + \frac{\partial^2 L}{\partial S^2}f^2\Bigg)e^{-\int B d\tau_1}d\tau_2\Bigg]e^{\int B d\tau_3}.\\
    \end{aligned}
\end{equation} 
It is tempting, at this point, to think that one can simply set $f = 0$, because the first variation of the action vanishes along a stationary path. We must remember, however, that this condition only applies at the endpoints of the system's trajectory (that is, $f(t_f) = f(0) = 0$, but for some abitrary time $t$, $f(t) \neq 0 $). We therefore need to be careful in handling the terms proportional to $f$ in Eq. (\ref{eq:f'=}). At some arbitrary time $t$, Eq. (\ref{eq:f'=}) can be written in the form
\begin{equation}
    f(t)e^{-\int_0^t B(\tau_1)d\tau_1} = \int_0^t\left(\frac{\partial L}{\partial x_i}y_i + \frac{\partial L}{\partial \dot{x}_i}\dot{y}_i\right)e^{-\int_0^{\tau_2} B(\tau_1)d\tau_1}d\tau_2.
\end{equation}
Integration of the RHS by parts produces
\begin{equation}
\begin{aligned}
\label{eq:fe_bound}
    f(t)e^{-\int_0^t B(\tau_1)d\tau_1} = \int_0^{t}\bigg(\frac{\partial L}{\partial x_i} & - \frac{d}{dt}\frac{\partial L}{\partial \dot{x}_i} + \frac{\partial L}{\partial S}\frac{\partial L}{\partial \dot{x}_i}\bigg)y_i e^{-\int_0^{\tau_2} B(\tau_1)d\tau_1}d\tau_2\\
    & + \left[e^{-\int_0^{\tau_2}B(\tau_1)d\tau_1}\frac{\partial L}{\partial \dot{x}}y\right]^{t}_0.\\
\end{aligned}
\end{equation}
Because the Euler-Lagrange equation vanishes at any time $t$, the first term on the RHS of Eq. (\ref{eq:fe_bound}) vanishes. Eq. (\ref{eq:fe_bound}) then reduces to
\begin{equation}
    f(t)e^{-\int_0^t B(\tau_1)d\tau_1} = e^{-\int_0^{t}B(\tau_1)d\tau_1}\left[\frac{\partial L}{\partial \dot{x}_i}y_i\right]\bigg|_{\tau_2 = t} - e^{-\int_0^{0}B(\tau_1)d\tau_1}\left[\frac{\partial L}{\partial \dot{x}_i}y_i\right]\bigg|_{\tau_2 = 0}.
\end{equation}
Because $y_i(0) = 0$, it follows that
\begin{equation}
\label{eq:f(s)=}
    f(t) = \frac{\partial L}{\partial \dot{x}_i}y_i,
\end{equation}
where $\frac{\partial L}{\partial \dot{x}_i}y_i$ is evaluated at $\tau_1 = t$. Insertion of Eq. (\ref{eq:f(s)=}) into Eq. (\ref{eq:f'=}) produces
\begin{equation}
\label{eq:f'=ver2}
\begin{aligned}
    f' = e^{\int_0^t B d\tau_3}\int_0^t & \bigg(\frac{\partial^2 L}{\partial \dot{x}_i \dot{x}_j}\dot{y}_i\dot{y}_j + 2\left[\frac{\partial^2 L}{\partial x_i\partial\dot{x}_j} + \frac{\partial^2 L}{\partial S\partial \dot{x}_i}\frac{\partial L}{\partial \dot{x}_j}\right]y_i \dot{y}_j + \frac{\partial^2 L}{\partial x_i \partial x_j}y_i y_j\\
    & + 2\frac{\partial^2 L}{\partial x_i\partial S}\frac{\partial L}{\partial \dot{x}_j}y_jy_i + \frac{\partial^2 L}{\partial S^2}\frac{\partial L}{\partial \dot{x}_i}\frac{\partial L}{\partial \dot{x}_j}y_iy_j\bigg)e^{-\int_0^{\tau_2} B d\tau_1}d\tau_2.\\
\end{aligned}
\end{equation}
Eq. (\ref{eq:f'=ver2}) is the general form of the second variational derivative of the Herglotz action. Note that, if the Lagrangian does not depend on the action, then Eq. (\ref{eq:f'=ver2}) reduces to
\begin{equation}
\label{eq:f'=consv}
    f' = \int_0^t \left(\frac{\partial^2 L}{\partial \dot{x}_i \dot{x}_j}\dot{y}_i\dot{y}_j + 2\frac{\partial^2 L}{\partial x_i\partial\dot{x}_j}y_i\dot{y}_j + \frac{\partial^2 L}{\partial x_i \partial x_j}y_i y_j\right) d\tau_1,
\end{equation}
which is the second variation of the action for a conservative system.

In Sec. \ref{sec:3}, I show, following Ref. \citenum{Gray_Taylor_2007}, that the second variation of the Herglotz action vanishes for a particular variation when the system's trajectory terminates at a kinetic focus. I then show that the Herglotz action is a saddle point when the system's trajectory terminates after a kinetic focus.

\section{The Herglotz action and kinetic foci}
\label{sec:3}
In this section I examine the necessary and sufficient conditions for the second variation of the Herglotz action to vanish, for the first time, at the first kinetic focus on a given trajectory. The sufficient condition in this case is just the same as the sufficient condition for conservative systems proved in Ref. \citenum{Gray_Taylor_2007}. In that paper the authors showed that, given a stationary path (call it $x_0$), there is a point in time ($t_{\rm KF}$) at which the second variation of the action along $x_0$ vanishes, this being the point at which a nearby stationary path (call it $x_1$) coalesces with $x_0$ in the limit that their initial velocities are made equal. The argument in Ref. \citenum{Gray_Taylor_2007} makes no reference to the specific form of the action, only requiring that $\delta^2 S_0 > 0$ for $t < t_{\rm KF}$. As long as $\delta^2 S_{0} > 0$ for $t < t_{\rm KF}$, where in this case $S$ is the Herglotz action, then by the same logic the sufficient condition for $\delta^2 S$ to vanish is that the given stationary path ends at $t_{\rm KF}$.

The necessary condition will require a little more work, but this can be done along the same lines as Ref. \citenum{Gray_Taylor_2007}, from which I adapt the following argument. Let the action of $x_0$ be $S_0$, and let the action along $x_1$ be $S_1$. In this case $S_0$ can be expanded in the series 
\begin{equation}
\label{eq:S0=S1}
    S_0 = S_1 + \delta S_1 + \delta^2 S_1 + \delta^3 S_1 + ...
\end{equation}
and $S_1$ can be expanded in the series
\begin{equation}
\label{eq:S1=S0}
    S_1 = S_0 + \delta S_0 + \delta^2 S_0 + \delta^3 S_0 + ...
\end{equation}
In Eqs. (\ref{eq:S0=S1}) and (\ref{eq:S1=S0}) I have defined
\begin{equation}
    \delta^{(n)}S_m = \left(-1\right)^{nm} \frac{\epsilon^n}{n!} \left[\frac{d^n}{d\epsilon^n}S\right]\bigg |_{\epsilon \rightarrow 0}
\end{equation}
with $m = 0, 1$. The factor of $(-1)^{nm}$ in the definition is due to the fact that $x_{1,i} = x_{0,i} + \epsilon y_i$, while the reference path is $x_{0,i} = x_{1,i} - \epsilon y_i$. That is, to get from $x_{0,i}$ to $x_{1,i}$, one must vary $x_{0,i}$ in the ``positive direction" away from $x_{0,i}$, while to get from $x_{1,i}$ to $x_{0,i}$ requires a variation in the ``negative direction" away from $x_{1,i}$. This has the effect of inducing a negative sign in $\delta^3 S_1$, as we will see below (see also Ref. \citenum{Gray_Taylor_2007} for a discussion of this point).

If we subtract Eq. (\ref{eq:S0=S1}) from Eq. (\ref{eq:S1=S0}), we get
\begin{equation}
\label{eq:S_diff}
    S_1 - S_0 = \left(S_0 - S_1\right) + \left(\delta S_0 - \delta S_1\right) + \left(\delta^2 S_0 - \delta^2 S_1\right) + \left(\delta^3 S_0 - \delta^3 S_1\right) + ...
\end{equation}
Because both paths are stationary, $\delta S_0 = \delta S_1 = 0
$. Then Eq. (\ref{eq:S_diff}) can be written in the form
\begin{equation}
\label{eq:S1-S0=}
    S_1 - S_0 = \frac{1}{2}\left(\delta^2 S_0 - \delta^2 S_1\right) + \frac{1}{2}\left(\delta^3 S_0 - \delta^3 S_1\right) + ...
\end{equation}
From Eq. (\ref{eq:f'=ver2}),
\begin{equation}
\label{eq:d^2S_1/de^2=}
    \frac{d^2 S}{d\epsilon^2}\bigg |_{\epsilon \rightarrow 0} = e^{\int_0^t B d\tau_3}\int^t_0\left(P_{ij}\dot{y}_i \dot{y}_j + 2Q_{ij}y_i \dot{y}_j + r_{ij}y_i y_j\right)e^{-\int^{\tau_2}_0 Bd\tau_1}d\tau_2,
\end{equation}
where I have defined
\begin{equation}
    \label{eq:P_def}
    P_{ij} := \frac{\partial^2 L}{\partial \dot{x}_i \partial \dot{x}_j},
\end{equation}
\begin{equation}
    \label{eq:Q_def}
    Q_{ij} := \frac{\partial^2 L}{\partial x_i\partial\dot{x}_j} + \frac{\partial^2 L}{\partial S\partial \dot{x}_i}\frac{\partial L}{\partial \dot{x}_j},
\end{equation}
and
\begin{equation}
\label{eq:r_def}
    r_{ij} := \frac{\partial^2 L}{\partial x_i \partial x_j} + 2\frac{\partial^2 L}{\partial x_i\partial S}\frac{\partial L}{\partial \dot{x}_j} + \frac{\partial^2 L}{\partial S^2}\frac{\partial L}{\partial \dot{x}_i}\frac{\partial L}{\partial \dot{x}_j}.
\end{equation}
Each of these functions can be expanded around $x_0$, so that they take the forms
\begin{equation}
    P_{ij}(x_1) \approx P_{ij}(x_0) + \frac{\partial P_{ij}(x_0)}{\partial x_k}\left(x^{(1)}_k - x^{(0)}_k\right) = P_{ij}(x_0) + \epsilon\frac{\partial P_{ij}(x_0)}{\partial x_k}y_k,
\end{equation}
\begin{equation}
    Q_{ij}(x_1) \approx Q_{ij}(x_0) + \frac{\partial Q_{ij}(x_0)}{\partial x_k}\left(x^{(1)}_k - x^{(0)}_k\right) = Q_{ij}(x_0) + \epsilon\frac{\partial Q_{ij}(x_0)}{\partial x_k}y_k,
\end{equation}
and
\begin{equation}
    r_{ij}(x_1) \approx r_{ij}(x_0) + \frac{\partial r_{ij}(x_0)}{\partial x_k}\left(x^{(1)}_k - x^{(0)}_k\right) = r_{ij}(x_0) + \epsilon\frac{\partial r_{ij}(x_0)}{\partial x_k}y_k
\end{equation}
to first order in $\epsilon$. From these expansions, we can see that the secondary Lagrangian, defined in Ref. \citenum{Morse} as
\begin{equation}
    \label{eq:Omega_def}
    2\Omega := P_{ij}\dot{y}_i \dot{y}_j + 2Q_{ij}y_i \dot{y}_j + r_{ij}y_i y_j,
\end{equation}
can be written
\begin{equation}
    2\Omega(x_1) \approx 2\Omega(x_0) + 2\frac{\partial \Omega(x_0)}{\partial x_k}\left(x^{(1)}_k - x^{(0)}_k\right) = 2\Omega(x_0) + 2\epsilon\frac{\partial \Omega(x_0)}{\partial x_k}y_k.
\end{equation}
Strictly speaking, it isn't necessary to introduce the secondary Lagrangian. I have done it here in order to reduce clutter in the following equations. Like the secondary Lagrangian, the exponential integrals can be expanded to linear order, giving
\begin{equation}
    e^{\int B^{(1)} d\tau} \approx e^{\int B^{(0)}d\tau}e^{ \int\epsilon\frac{\partial B(x_0)}{\partial x_k}y_k d\tau} = e^{\int B^{(0)} d\tau}\left(1 + \epsilon\int\frac{\partial B(x_0)}{\partial x_k}y_k d\tau\right).
\end{equation}
Therefore, after a bit of simplification, the first term on the RHS of Eq. (\ref{eq:S1-S0=}) becomes
\begin{equation}
\begin{aligned}
\label{eq:delta^2S_diff}
    \frac{1}{2}\left(\delta^2 S_0 - \delta^2 S_1\right) & = -\frac{\epsilon^3}{4} e^{\int B d\tau_3}\bigg\{\int_0^t \left[2\frac{\partial \Omega}{\partial x_k}y_k - 2\Omega\left(\int_0^{\tau_2}\frac{\partial B}{\partial x_k}y_k d\tau_1\right) \right]e^{-\int_0^{\tau_2} B d\tau_1} d\tau_2\\
    & + \int^t_0\frac{\partial B}{\partial x_k}y_k d\tau_3\int_0^t 2\Omega e^{-\int_0^{\tau_2} B d\tau_1} d\tau_2\bigg\} + O(\epsilon^4),\\
\end{aligned}
\end{equation}
where $\Omega$, $\frac{\partial\Omega}{\partial x_k}$, $B$, and $\frac{\partial B}{\partial x_k}$ are evaluated on $x_0$.

The key point to keep in mind here is that Eq. (\ref{eq:delta^2S_diff}) is $O(\epsilon^3)$ to lowest order, which ultimately means (see below) that the second variation of the action will vanish when the system terminates at a kinetic focus. It also means that, in calculating $S_0 - S_1$, we can't ignore the $\delta^3S_0 - \delta^3S_1$ term in Eq. (\ref{eq:S1-S0=}). Calculating the third functional derivative of $S$ (or, equivalently, the second functional derivative of $f$) leads to
\begin{equation}
    \frac{df''}{dt} = \tilde{A}'' + B''f + 2B'f' + Bf'',
\end{equation}
which is solved by
\begin{equation}
    f'' = e^{\int B d\tau_3}\left[\int_0^t \left(\tilde{A}'' + B''f + 2B'f'\right)e^{-\int_0^{\tau_2} B d\tau_1} d\tau_2\right].
\end{equation}
From Eq. (\ref{eq:f(s)=}), we have
\begin{equation}
\label{eq:f''=}
    f'' = e^{\int B d\tau_3}\left[\int_0^t \left(\tilde{A}'' + B''\frac{\partial L}{\partial \dot{x}_i}y_i + 2B'f'\right)e^{-\int_0^{\tau_2} B d\tau_1} d\tau_2\right].
\end{equation}
As before, one can Taylor expand the RHS of this equation around $x_0$. This is a messier calculation than what I have shown above, so the details  are sketched out in Appendix \ref{app:B}. The main point is that, to lowest order, and with $B$, $\Omega$, $\frac{\partial \Omega}{\partial x_k}$, and $\frac{\partial B}{\partial x_k}$ evaluated on $x_0$,
\begin{equation}
\begin{aligned}
\label{eq:S1-S0=full}
    S_1 - S_0 & = \frac{1}{2}\left(\delta^2 S_0 - \delta^2 S_1\right) + \frac{1}{2}\left(\delta^3 S_0 - \delta^3 S_1\right)\\
    & = \frac{\epsilon^2}{4}\left(f'_{(0)} - f'_{(1)}\right) + \frac{\epsilon^3}{12}\left(f''_{(0)} + f''_{(1)}\right)\\
    & = -\frac{\epsilon^3}{4}\bigg\{ e^{\int B d\tau_3}\int_0^t \left[2\frac{\partial \Omega}{\partial x_k}y_k - 2\Omega\int_0^{\tau_2}\frac{\partial B}{\partial x_k}y_k d\tau_1 \right]e^{-\int_0^{\tau_2} B d\tau_1} d\tau_2\\
    & + e^{\int B d\tau_3}\int_0^t\frac{\partial B}{\partial x_k}y_k d\tau_3\int_0^t 2\Omega e^{-\int_0^{\tau_2} B d\tau_1} d\tau_2\bigg\} \\
    & + \frac{\epsilon^3}{6}\bigg\{e^{\int B d\tau_3}\bigg[\int_0^t \bigg(\tilde{A}'' + B''\frac{\partial L}{\partial \dot{x}_i}y_i + 2B'f'\bigg)e^{-\int_0^{\tau_2}Bd\tau_1}d\tau_2\bigg]\bigg\}.\\
\end{aligned}
\end{equation}
This equation tells us that $S_1 - S_0 \sim O(\epsilon^3)$ to lowest order when the stationary path terminates near a kinetic focus (conjugate point). This has two consequences. First, and as mentioned above, $\delta^2 S_0 = 0$ when $t_f = t_{\rm KF}$. Second, the action of the stationary path $x_0$ is a saddle point when the stationary path terminates a short time after $t_{\rm KF}$.

That $\delta^2 S_0 = 0$ when $t_f = t_{\rm KF}$ follows from a comparison of Eqs. (\ref{eq:S1-S0=}) and (\ref{eq:S1=S0}). Subtraction of $S_0$ from both sides of Eq. (\ref{eq:S1=S0}) produces
\begin{equation}
    S_1 - S_0 = \delta^2 S_0 + \delta^3 S_0 + ...
\end{equation}
Because $S_1 - S_0$ in Eq. (\ref{eq:S1-S0=}) has the form $-\frac{\epsilon^3}{4}\{...\} + \frac{\epsilon^3}{6}\{...\}$, and $\delta^3 S_0 = -\frac{\epsilon^3}{12}\frac{d^3 S_0}{d\epsilon^3}$, it follows that $\delta^2 S_0$ must also be $O(\epsilon^3)$ if $t_f$ is near $t_{\rm KF}$. The difference $S_1 - S_0$, which is $O(\epsilon^3)$ when $t_f$ is near $t_{\rm KF}$, vanishes when $\epsilon \rightarrow 0$ (that is, when $x_1$ and $x_0$ coalesce). Therefore $\delta^2 S_0$, which is also $O(\epsilon^3)$, must also vanish when $\epsilon \rightarrow 0$. If the signs and relative magnitudes of the quantities in the curly brackets are such that $S_1 - S_0 > 0$ for a variation that terminates before $t_{\rm KF}$, then $S_0$ will be minimized for $t < t_{\rm KF}$. We have already seen that $S_1 - S_0 = 0$ when $\epsilon \rightarrow 0$, which occurs when the varied path terminates at a kinetic focus. When the varied path terminates after a kinetic focus, then $\epsilon y_k$ will be negative, ensuring that the varied path $x_1$ has an action which is smaller than the action of $x_0$ (see Fig. 12 of Ref. \citenum{Gray_Taylor_2007}). $\delta^2 S_0$ will then be negative for a variation away from $x_0$ toward $x_1$, but will remain positive for all other variations. $S_0$ is therefore a saddle point when $x_0$ terminates slightly later than $t_{\rm KF}$. As an example, consider the Herglotz Lagrangian
\begin{equation}
\label{eq:L=Herglotz}
    L = \frac{1}{2}m\dot{x}^2 - U(t, x) + g(t)S.
\end{equation}
Many dissipative systems studied in the literature are governed by Lagrangians of the form shown in Eq. (\ref{eq:L=Herglotz}),\cite{Sloan_Cosmology_2021, Lazo_et_al_JMP_2018, Lazo_et_al_PRD_2017, Georgieva_Guenther_2005, Georgieva_Guenther_Bodurov_JMP_2003, Paiva_Lazo_Zanchin_arXiv_2021, Zhang_Tian_PLA_2019, Tian_Zhang_Adiabatic_2020} as are the two examples that I study in sections \ref{subsec:SHO_a} and \ref{subsec:SHO_g}, so it is worthwhile to consider this special case. In this case all partial derivatives of $B$ with respect to $x$ vanish, $2\frac{\partial \Omega(x_0)}{\partial x} = -U'''(t, x_0)y^3$, and $\tilde{A}'' = -U'''(t, x_0)y^3$. Then Eq. (\ref{eq:S1-S0=}) becomes
\begin{equation}
\label{eq:S1-S0=simplified}
    S_1 - S_0 = \frac{\epsilon^3}{12}e^{\int_0^t B^{(0)}d\tau_3}\int_0^t U'''(t, x_0)y^3 e^{-\int_0^{\tau_2} B^{(0)} d\tau_1} d\tau_2.
\end{equation}
If $U'''(t, x) > 0$ up to time $t$, then the integrand in Eq. (\ref{eq:S1-S0=simplified}) will be positive, and the sign of $S_1 - S_0$ will only change when $\epsilon y$ changes sign, just beyond $t_{\rm KF}$. Writing $S_1 - S_0$ in this form not only makes the behavior of the action clearer, but it also makes it easier to see the relation between the action of a non-conservative, one dimensional system described by Eq. (\ref{eq:L=Herglotz}) and the action of a conservative one-dimensional system studied in Ref. \citenum{Gray_Taylor_2007}. $B$ vanishes when $g(t) = 0$, in which case the results of this section reduce to those of Ref. \citenum{Gray_Taylor_2007} (assuming a time-independent potential energy function).

To summarize the results of this section, the second variation of the Herglotz action of a non-conservative system vanishes for the first time when the trajectory of the system terminates at that trajectory's first kinetic focus. The Herglotz action is a saddle point ($\delta^2 S < 0$ for at least one variation, $\delta^2 S > 0$ for all others) if the trajectory terminates after the first kinetic focus. These statements generalize the results obtained in Refs. \citenum{Gray_Taylor_2007} and \citenum{Gray_Poisson_AJP_2011} for conservative systems, showing that the analysis of $\delta^2 S$ for a non-conservative system described by the Herglotz action can proceed along the same lines as the analysis of $\delta^2 S$ for a conservative system, only with somewhat more complicated equations. One convenient method for calculating the sign of $\delta^2 S$ for conservative systems is described in Refs. \citenum{Hussein_et_al_1980} and \citenum{Levit_Smilansky_AnnPhys_1977}. In this method, the equation for $\delta^2 S$ (Eq. \ref{eq:f'=consv} here) is transformed into a linear eigenvalue equation, with the signs of the eigenvalues reflecting the sign of $\delta^2 S$ at a particular value of $t_f$. I will apply this method to Eq. (\ref{eq:f'=}) in the Sec. \ref{sec:eig_analysis} and demonstrate its use with two examples.

\section{Eigenvalue analysis of second variation of Herglotz action}
\label{sec:eig_analysis}
First, because we're interested in the sign of $f'$, we needn't worry about the $e^{\int B d\tau_2}$ term outside the integral. Because the exponential is positive-definite along the system's entire trajectory, it will affect the amplitude of $f'$ but not the location of its roots, so it won't have any effect on when the sign of $f'$ changes (assuming, as I do throughout this paper, that $B(t)$ is a real-valued function). We can therefore define a new quantity
\begin{equation}
\label{eq:d2A_def}
    \delta^2 \cA := f' e^{-\int B d\tau_3}
\end{equation}
and determine how the sign of this quantity changes with $t$. We want to determine the sign of
\begin{equation}
\label{eq:d2A=}
\begin{aligned}
    \delta^2 \cA = \int_0^t\bigg(\frac{\partial^2 L}{\partial \dot{x}_i \dot{x}_j}\dot{y}_i\dot{y}_j & + 2\left[\frac{\partial^2 L}{\partial x_i\partial\dot{x}_j}y_i + \frac{\partial^2 L}{\partial S\partial \dot{x}_i}\frac{\partial L}{\partial \dot{x}_j}y_i\right]\dot{y}_j + \frac{\partial^2 L}{\partial x_i \partial x_j}y_i y_j\\
    & + 2\frac{\partial^2 L}{\partial x_i\partial S}\frac{\partial L}{\partial \dot{x}_j}y_jy_i + \frac{\partial^2 L}{\partial S^2}\frac{\partial L}{\partial \dot{x}_i}\frac{\partial L}{\partial \dot{x}_j}y_iy_j\bigg)e^{-\int_0^{\tau_2} B d\tau_1}d\tau_2.\\
\end{aligned}
\end{equation}
Following Refs. \citenum{Hussein_et_al_1980} and \citenum{Levit_Smilansky_AnnPhys_1977}, I will define
\begin{equation}
    \label{eq:R_def}
    R_{ij} := r_{ij} + BQ_{ji}  + \frac{B^2}{4}P_{ij} - \frac{1}{2}\frac{dB}{dt}P_{ij} - \frac{B}{2}\frac{dP_{ij}}{dt},
\end{equation}
and
\begin{equation}
\label{eq:Lambda_def}
    \Lambda_{ij} := -\frac{d}{dt}\left(P_{ij}\frac{d}{dt} + Q_{ji}\right) + \left(Q_{ij}\frac{d}{dt} + R_{ij}\right),
\end{equation}
where $P_{ij}$, $Q_{ij}$ and $r_{ij}$ are defined in Eqs. (\ref{eq:P_def}-\ref{eq:r_def}). One can then show (see Appendix \ref{app:A}), that Eq. (\ref{eq:d2A=}) takes the form
\begin{equation}
\label{eq:d2A=Lambda}
    \delta^2 \cA = \int_0^t \eta_i \Lambda_{ij} \eta_j \hspace{0.5mm}d\tau_2
\end{equation}
where the $\{\eta_i\}$ functions are defined to be
\begin{equation}
\label{eq:eta_def}
    \eta_i := y_i e^{-\frac{1}{2}\int_0^{t}Bd\tau_1}.
\end{equation}
By defining the inner product
\begin{equation}
    \left(\eta_i, \eta_j\right) := \int \eta_i \eta_j\hspace{0.5mm} d\tau,
\end{equation}
and then expanding the $\{\eta_i\}$ functions in an orthonormal basis $\{b_{in}\}$
\begin{equation}
    \eta_i = \sum_n \beta_n b_{in},
\end{equation}
with
\begin{equation}
    \int b_n b_m d\tau = \delta_{nm},
\end{equation}
then Eq. (\ref{eq:d2A=}) takes the form
\begin{equation}
\delta^2 \cA = \sum_{n, m} A_{nm}\beta_{n}\beta_{m},
\end{equation}
where
\begin{equation}
    A_{nm} = \left(b_n, \Lambda b_m\right) = \int b_{in} \Lambda_{ij} b_{jm} d\tau.
\end{equation}
If the basis diagonalizes $A_{nm}$, then
\begin{equation}
\label{eq:Lambda=lambda}
    \Lambda_{ij} b_{jn} = \lambda_n b_{jn}.
\end{equation}
We have now reduced the problem of determining the sign of $\delta^2 \cA$ to a linear eigenvalue problem. If all the eigenvalues are positive at time $t$, then the action is minimized up to that time. If one of the eigenvalues ($\lambda_n$) vanishes at time $t$, then the action along the trajectory of the $n^{\rm th}$ variation vanishes, and $t$ is a kinetic focus. If one of the eigenvalues ($\lambda_n$) is negative at time $t$, then the action is maximized along the trajectory of the $n^{\rm th}$ variation.\cite{Papastavridis_1983} Determining the sign of the Herglotz action is then just a matter of working out the eigenvalue spectrum $\{\lambda_n\}$ from Eq. (\ref{eq:Lambda=lambda}). To illustrate the use of this technique, I will make the simplifying assumption that the Herglotz Lagrangian takes the form shown in Eq. (\ref{eq:L=Herglotz}). Then Eq. (\ref{eq:Lambda=lambda}) becomes
\begin{equation}
\label{eq:Hamiltonian_analogue}
    \left(-P\frac{d^2}{dt^2} + R\right)b_n = \lambda_n b_n,
\end{equation}
where
\begin{equation}
\label{eq:P_1D}
    P = \frac{\partial^2 L}{\partial \dot{x}^2} = m,
\end{equation}
and
\begin{equation}
\label{eq:R_1D}
    R = \frac{\partial^2 L}{\partial x^2} + \frac{1}{4}\left(g^2 - 2\frac{dg}{dt}\right)\frac{\partial^2 L}{\partial \dot{x}^2} = -\frac{\partial^2 U}{\partial x^2} + \frac{m}{4}\left(g^2 - 2\frac{dg}{dt}\right).
\end{equation}
Eq. (\ref{eq:Hamiltonian_analogue}) is, as described in Ref. \citenum{Hussein_et_al_1980}, directly analogous to the eigenvalue equation for a particle in a box with an effective mass $m = \frac{\hbar^2}{2P}$ moving in an effective potential
\begin{equation}
    \mathcal{U} = 
    \begin{cases}
    R & \text{if}\ 0 < t < t_f,\\
    \vspace{1mm}
    \infty & \text{if}\ t = 0 \hspace{1mm} {\rm or} \hspace{1mm} t = t_f ,\\
    \end{cases}
\end{equation}
with $t$ playing the role that $x$ ordinarily plays in time-independent perturbation theory. Inserting Eqs. (\ref{eq:P_1D}) and (\ref{eq:R_1D}) into Eq. (\ref{eq:Hamiltonian_analogue}), we find
\begin{equation}
\label{eq:short_eigenvalue_equation}
    \left(-\frac{d^2}{dt^2} - \frac{\partial^2 \tilde{U}}{\partial x^2} + \frac{g^2}{4} - \frac{1}{2}\frac{dg}{dt}\right)b_n = \frac{\lambda_n}{m} b_n,
\end{equation}
where $\tilde{U} := \frac{U}{m}$. In general, $\tilde{U}$ is a function of the solution $x(t)$ of the equation of motion given by Eq. (\ref{eq:EoM}). If this solution is known, then one can write $\tilde{U}$ explicitly as a function of time, and then attempt to solve Eq. (\ref{eq:short_eigenvalue_equation}). In the special case that $\frac{\partial^2\tilde{U}}{\partial x^2} = F(t) + C$, where $F(t)$ is a function only of time and $C$ is a constant, one can solve the eigenvalue spectrum of the non-conservative system without solving the equation of motion beforehand. Eq. (\ref{eq:short_eigenvalue_equation}) can be solved in a number of ways that are familiar from non-relativistic quantum mechanics, such as perturbation theory, the variational method, the WKB method, and others.\cite{Hussein_et_al_1980, Schumacher_Westmoreland_2010, Shankar} In Secs. \ref{subsec:SHO_a} and \ref{subsec:SHO_g} I will consider two examples: a harmonic oscillator with time-independent damping, and a harmonic oscillator with time-dependent damping (and time-dependent frequency).

\subsection{Harmonic oscillator with time-independent damping}
\label{subsec:SHO_a}
For an oscillator with time-independent damping, let $g(t) = -a$ and $\tilde{U} = \frac{1}{2}\omega_0^2 x^2$, where $a$ and $\omega_0$ are real-valued constants. In this case Eq. (\ref{eq:short_eigenvalue_equation}) reduces to
\begin{equation}
    \left(-\frac{d^2}{dt^2} - \omega_d^2\right)b_n = \frac{\lambda_n}{m}b_n,
\end{equation}
where I have defined $\omega_d^2 := \omega_0^2 - \frac{a^2}{4}$. From Refs. \citenum{Gray_Taylor_2007} and \citenum{Hussein_et_al_1980}, we know that the action eigenvalue spectrum of an undamped harmonic oscillator is
\begin{equation}
    \frac{\lambda_n^c}{m} = \left(\frac{n\omega}{2}\right)^2 - \omega_0^2,
\end{equation}
where $\omega := \frac{2\pi}{t}$, so the eigenvalue spectrum of the damped oscillator must be
\begin{equation}
\label{eq:SHO_action_eigv_spec}
    \frac{\lambda_n}{m} = \left(\frac{n\omega}{2}\right)^2 - \omega_0^2 + \frac{a^2}{4}.
\end{equation}
This leads to a familiar result. As is well known, the frequency of a harmonic oscillator subject to linear damping, with an equation of motion given by
\begin{equation}
    \ddot{x} + a\dot{x} + \omega_0^2 x = 0,
\end{equation}
is $\omega_{\rm d} = \omega_0\sqrt{1 - \left(\frac{a}{2\omega_0}\right)^2}$ (see Ref. \citenum{Fowles_Cassiday_2007}, for example). If $n = 1$, then the second variation of the action vanishes when
\begin{equation}
    0 = \left(\frac{\omega}{2}\right)^2 - \omega_0^2 + \frac{a^2}{4},
\end{equation}
or
\begin{equation}
    \omega = 2\omega_0\sqrt{1 - \left(\frac{a}{2\omega_0}\right)^2} = 2\omega_d.
\end{equation}
From Eq. (\ref{eq:kinetic_focus}), one can show that the first kinetic focus of a damped harmonic oscillator occurs when
\begin{equation}
\label{eq:t_KF1_DHO}
    t_{\rm KF,1} = \frac{2\pi}{\omega} = \frac{\pi}{\omega_d} = \frac{\pi}{\omega_0}\frac{1}{\sqrt{1 - \left(\frac{a}{2\omega_0}\right)^2}} = \frac{T_0}{2}\frac{1}{\sqrt{1 - \left(\frac{a}{2\omega_0}\right)^2}},
\end{equation}
which is just the half-period of the damped oscillator ($T_0$ is the undamped period).\footnote{This result was also obtained by G. Leitmann in Ref. \citenum{Leitmann_Remarks_1963}, with an action of the form $S = \int e^{2ct}\left[\dot{x}^2 - \frac{1}{2}\omega_0^2 x^2\right] dt$, where my $a$ equals his $2c$.} Eq. (\ref{eq:t_KF1_DHO}) also tells us that the time to reach the first kinetic focus diverges as the oscillator approaches critical damping (that is, when $a \rightarrow 2\omega_0$). In this case (or in the case of strong damping, for which $a > 2\omega_0$) the oscillations decay on a time scale that is shorter than $T_0/2$, so there are no kinetic foci, and the action is minimized along the entire trajectory of the oscillator.

\subsection{Harmonic oscillator with time-dependent damping}
\label{subsec:SHO_g}
Consider an equation of motion of the form
\begin{equation}
    \label{eq:TDHO_EoM}
    \ddot{x} + \gamma(t)\dot{x} + \omega^2(t)x = 0,
\end{equation}
where $\gamma(t)$ is the dissipation function and the frequency depends on time. This equation describes a time-dependent harmonic oscillator (TDHO), a system that has been used to model the motion of a non-relativistic charged particle in a time-dependent magnetic field \cite{Lewis_PRL_1967} as well as the normal mode dynamics of a scalar field in an expanding universe\cite{Robles-Perez_PLB_2017, Finelli_Vacca_Venturi_PRD_1998} (see also Ref. \citenum{Liu_Torres_Wang_AoP_2018, Bravetti_Cruz_Tapias_Ann_Phys_2017, de_Leon_et_al_Monatshefte_2022}).

Eq. (\ref{eq:TDHO_EoM}) can be derived from the Herglotz Lagrangian of an oscillator with time-dependent frequency and unit mass
\begin{equation}
    L = \frac{1}{2}\dot{x}^2 - \frac{1}{2}\omega^2(t)x^2 + g(t)S,
\end{equation}
where, in this case, $g(t) = -\gamma(t)$. Eq. (\ref{eq:short_eigenvalue_equation}) tells us that the action eigenvalue equation for the TDHO is
\begin{equation}
\label{eq:Herglotz_g(t)_eig}
    \left(-\frac{d^2}{dt^2} - \omega^2(t) + \frac{1}{4}\gamma^2(t) + \frac{1}{2}\dot{\gamma}(t)\right)b_n = \lambda_n b_n.
\end{equation}
To simplify the problem, I will assume that the damping function and the frequency vary slowly over the time interval $0 \leq t \leq t_f$. I will also assume that these functions can be Taylor expanded around $t = 0$, so that
\begin{equation}
    \label{eq:TDHO_freq_Taylor}
    \omega(t) = \omega(0) + t\frac{d\omega}{dt}\bigg|_{t = 0} + O(t^2),
\end{equation}
and
\begin{equation}
    \label{eq:TDHO_gamma_Taylor}
    \gamma(t) = \gamma(0) + t\frac{d\gamma}{dt}\bigg|_{t = 0} + O(t^2).
\end{equation}
Define $\omega_0 := \omega(0)$, $\omega_0' = \frac{d\omega}{dt}\big|_{t = 0}$, $\gamma_0 := \gamma(0)$, and $\gamma_0' = \frac{d\gamma}{dt}\big|_{t = 0}$. Then Eq. (\ref{eq:Herglotz_g(t)_eig}) takes the form
\begin{equation}
\label{eq:Eff_Schro}
    \left(-\frac{d^2}{dt^2} - \omega_0^2 + \frac{\gamma_0^2}{4} + \frac{\gamma_0'}{2} - 2\left(\omega_0\omega_0' - \frac{\gamma_0\gamma_0'}{4}\right)t - \left(\omega'^2_{0} - \frac{\gamma'^2_{0}}{4}\right)t^2\right)b_n = \lambda_n b_n.
\end{equation}
Here we have an effective Schr\"odinger equation for a particle of unit mass moving in an effective potential
\begin{equation}
    \mathcal{U} = 
    \begin{cases}
    - \omega_0^2 + \frac{\gamma_0^2}{4} + \frac{\gamma_0'}{2} - 2\left(\omega_0\omega_0' - \frac{\gamma_0\gamma_0'}{4}\right)t - \left(\omega'^2_{0} - \frac{\gamma'^2_{0}}{4}\right)t^2 & \text{if}\ 0 < t < t_f,\\
    \vspace{1mm}
    \infty & \text{if}\ t = 0 \hspace{1mm} {\rm or} \hspace{1mm} t = t_f.\\
    \end{cases}
\end{equation}
We can see that, depending on the initial values of the frequency and the damping function (as well as the initial values of their first derivatives), the depth of the effective potential can increase with time. Therefore, even if the system is critically damped at first (so that $\omega_0 = 2\gamma_0$), it may eventually develop a kinetic focus. This happens because, for a sufficiently deep effective potential, one or more of the action eigenvalues can be $\leq 0$ (the analogous situation in quantum mechanics would be to a particle in a potential well that is deep enough to admit energy eigenvalues that are $\leq 0$).

\begin{figure}[ht]
\centering
    \resizebox{\columnwidth}{!}{%
    \includegraphics[scale=1]{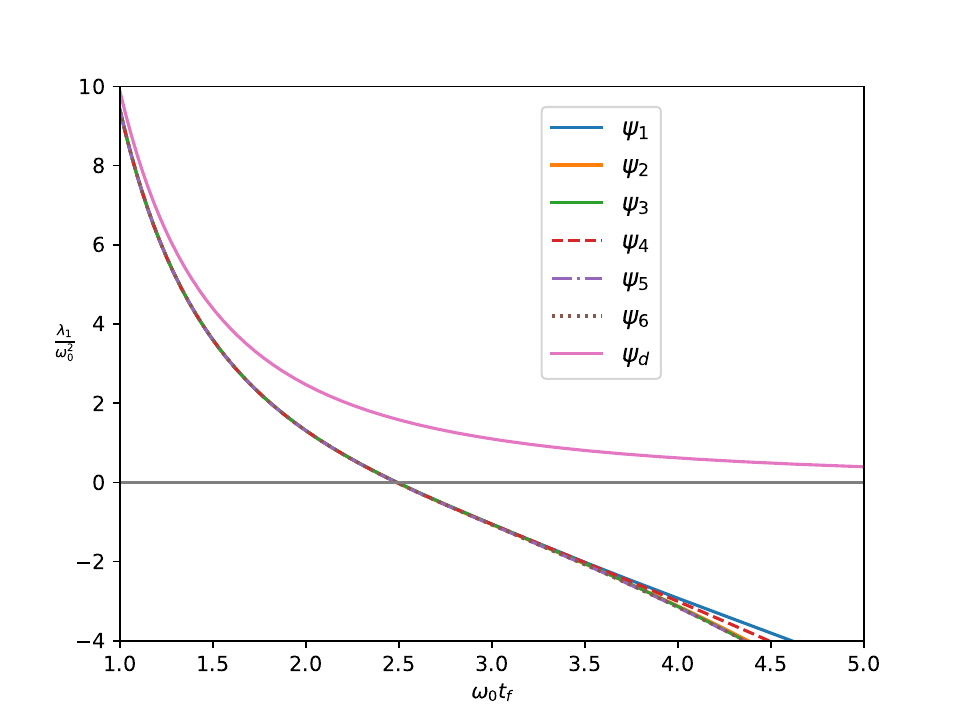}%
    }
\caption{Lowest action eigenvalue, normalized to $\omega_0^2$, versus $\omega_0 t_f$, where $t_f$ is the time at which the stationary path of the TDHO terminates. The solid pink curve corresponds to a time-independent damped harmonic oscillator, and the other curves correspond to the trial functions shown in Table \ref{tab:Trial_functions}.}
\label{fig:Ground_eig_var_est_python}
\end{figure}

\begin{table*}
    \centering
    \begin{tabular}{c|c}
        \hline
         TDHO trajectories & $v(t = 0)$\\
         \hline
         $x_0(t)$ & 1\\
         $x_1(t)$ & 1.05\\
         $x_2(t)$ & 1.10\\
         $x_3(t)$ & 1.15\\
         $x_4(t)$ & 1.20\\
         \hline
    \end{tabular}
    \caption{Trajectories used to generate Figs. \ref{fig:TDHO_x} and \ref{fig:Damped_HO_x}. The left column lists the trajectories with their corresponding label subscripts, and the right column lists the initial values of their velocities.}
    \label{tab:Initial_conditions}
\end{table*}

\begin{table*}
    \centering
    \begin{tabular}{c|c|c}
        \hline
         Variational parameters & Trial function & Legend label\\
         \hline
         $\alpha$ & $N\left(1 + \alpha t\right){\rm sin}\left(\frac{\pi t}{t_f}\right)$ & $\psi_1$\\
         $\alpha$ & $Ne^{\alpha t}{\rm sin}\left(\frac{\pi t}{t_f}\right)$ & $\psi_2$\\
         $\alpha$ & $Ne^{\alpha t^2}{\rm sin}\left(\frac{\pi t}{t_f}\right)$ & $\psi_3$\\
         $\alpha$, $\beta$ & $N\left(\alpha + \beta t\right){\rm sin}\left(\frac{\pi t}{t_f}\right)$ & $\psi_4$\\
         $\alpha$, $\beta$, $\gamma$ & $N\left(\alpha + \beta t + \gamma t^2\right){\rm sin}\left(\frac{\pi t}{t_f}\right)$ & $\psi_5$\\
         $\alpha$, $\beta$, $\gamma$, $\delta$ & $N\left(\alpha + \beta t + \gamma t^2 + \delta t^3\right){\rm sin}\left(\frac{\pi t}{t_f}\right)$ & $\psi_6$\\
         \hline
    \end{tabular}
    \caption{Trial functions for the variational calculation of the lowest action eigenvalue of the TDHO. $N$ is a normalization constant and $t_f$ is the time at which the stationary path terminates.}
    \label{tab:Trial_functions}
\end{table*}

To investigate this possibility, I will estimate the lowest eigenvalue of the action spectrum, as a function of the final time $t_f$, via the variational method. In standard quantum mechanics, it is possible to estimate the ground state energy $E_0$ of a system with a given Hamiltonian $H$ by calculating the quantity
\begin{equation}
    E(\alpha) = \frac{\bra{\psi_{\alpha}}H\ket{\psi_{\alpha}}}{\bra{\psi_{\alpha}}\ket{\psi_{\alpha}}},
\end{equation}
and demanding that $\alpha$ take its minimum value (in this equation $\ket{\psi_{\alpha}}$ is a state that depends on the variational parameter $\alpha$). Then $E(\alpha_{\rm min})$ provides an upper bound on the ground state energy such that $E(\alpha_{\rm min}) \geq E_0$.\cite{Schumacher_Westmoreland_2010, Shankar} The accuracy of the upper bound $E(\alpha_{\rm min})$ depends on how closely the trial wavefunction $\psi_{\alpha}$ approximates the ground state wavefunction $\psi_0$ (although it needn't be too close; even a relatively crude trial wavefunction can still give a reasonable approximation to the ground state energy \cite{Schumacher_Westmoreland_2010, Shankar}). In our case, we want
\begin{equation}
\label{eq:lambda_1}
    \lambda_1(\alpha_{\rm min}) = \frac{\left(\psi_{\alpha}, \mathcal{H} \psi_{\alpha}\right)}{\left(\psi_{\alpha}, \psi_{\alpha}\right)}\bigg|_{\alpha = \alpha_{\rm min}},
\end{equation}
where the effective Hamiltonian is
\begin{equation}
    \mathcal{H} = -\frac{d^2}{dt^2} + \mathcal{U}.
\end{equation}
To compute $\lambda_1(\alpha_{\rm min})$, I wrote a \textsc{Python} script to minimize $\lambda_1(\alpha)$ by way of the \textsc{minimize} routine built into \textsc{SciPy}.\footnote{When I called this routine, I used the Sequential Least Squares Programming method. Documentation: \url{https://docs.scipy.org/doc/scipy/reference/generated/scipy.optimize.minimize.html}} I used several trial functions, shown in Table \ref{tab:Trial_functions}. The ${\rm sin}\left(\frac{\pi t}{t_f}\right)$ in each trial function is to ensure that the function vanishes at $t = 0$ and $t = t_f$. For the single parameter functions, I wanted to ensure that the trial function also decays rapidly in the direction of $t = 0$, hence the choice of a linear, exponential, or Gaussian function multiplying the sine. The multi-parameter functions are intended as a convergence test. The main difficulty with the variational method in quantum mechanics is that one can never be completely sure of how well it approximates the true ground state energy of a given system, and the same is true of the lowest order action eigenvalue here. Generally speaking, if one finds that the variational estimate does not change significantly as one uses increasingly complicated trial functions, then one can be fairly confident that the estimate is reliable.\cite{Schumacher_Westmoreland_2010, Shankar} The results of this analysis are shown in Fig. \ref{fig:Ground_eig_var_est_python}. To plot the results in a scale-independent fashion, I divided both sides of Eq. (\ref{eq:Eff_Schro}) by $\omega_0^2$ to obtain
\begin{equation}
\begin{aligned}
\label{eq:Scale_independent_Hamiltonian}
    \Bigg[-\frac{d^2}{d\left(\omega_0 t\right)^2} - 1 + \frac{\gamma_0^2}{4\omega_0^2} + \frac{\gamma_0'}{2\omega_0^2} & - 2\left(\frac{\omega_0'}{\omega_0^2} - \frac{\gamma_0\gamma_0'}{4\omega_0^3}\right)\left(\omega_0 t\right)\\
    & - \left(\frac{\omega'^2_{0}}{\omega_0^4} - \frac{\gamma'^2_{0}}{4\omega_0^4}\right)\left(\omega_0 t\right)^2\Bigg]b_n = \frac{\lambda_n}{\omega_0^2} b_n.\\
\end{aligned}
\end{equation}
and used the effective Hamiltonian shown on the LHS of Eq. (\ref{eq:Scale_independent_Hamiltonian}) in the variational calculation of Eq. (\ref{eq:lambda_1}). In Fig. \ref{fig:Ground_eig_var_est_python}, I plot $\frac{\lambda_1}{\omega_0^2}$ versus $\omega_0 t_f$, after (somewhat arbitrarily) setting $\gamma_0 = 2\omega_0$, $\omega_0' = \omega_0^2/2$, $\gamma_0' = \omega_0^2/10$, and making the replacements $t \rightarrow \omega_0 t$ and $t_f \rightarrow \omega_0 t_f$ in the trial functions of Table \ref{tab:Trial_functions}. The solid pink curve corresponds to a critically damped, time-independent harmonic oscillator. Each of the other curves corresponds to one of the trial functions shown in Table \ref{tab:Trial_functions}, labeled accordingly. From the figure, we can see that value of $\frac{\lambda_1}{\omega_0^2}$ for a TDHO differs noticeably from that of the time-independent harmonic oscillator, even down to $\omega_0 t_f \sim 1$. Near $\omega_0 t_f \approx 2.5$, the eigenvalue curve passes through zero, signaling the appearance of a kinetic focus. We can also see that all of the trial functions are in good agreement up to $\omega_0 t_f \approx 3.5$, showing that the prediction of a kinetic focus at $\omega_0 t_f \approx 2.5$ is robust, given the set of parameters I have chosen.

\begin{figure}[ht]
\centering
    \resizebox{\columnwidth}{!}{%
    \includegraphics[scale=1]{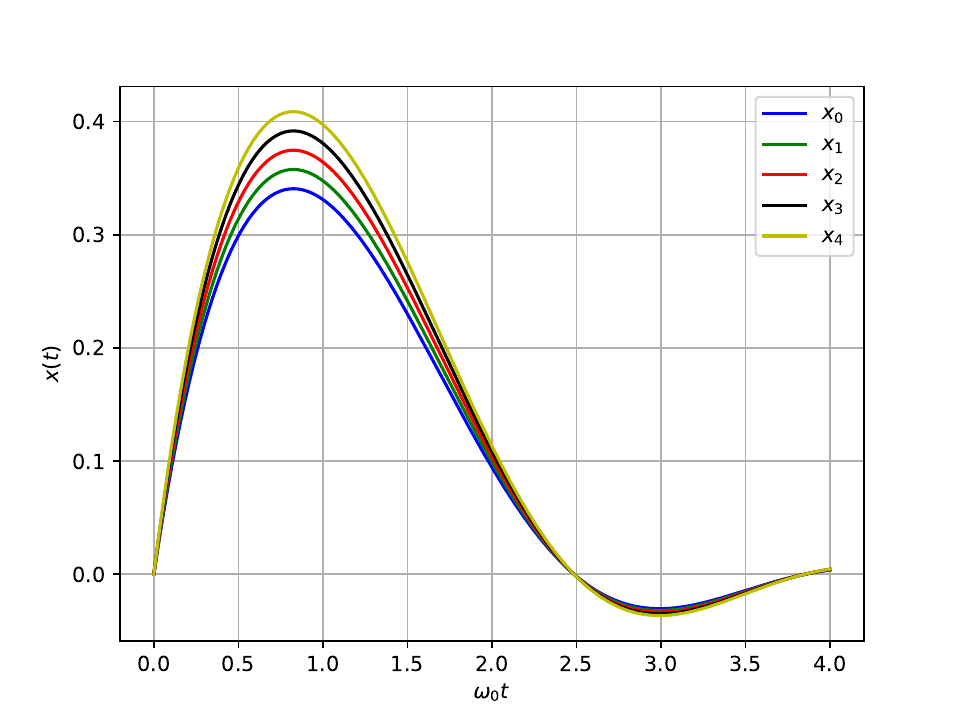}%
    }
\caption{Numerically obtained trajectories of a critically damped TDHO with slightly different initial velocities. See Table \ref{tab:Initial_conditions} for a list of these initial velocities. Here the displacement $x(t)$ is reported in arbitrary units.}
\label{fig:TDHO_x}
\end{figure}

\begin{figure}[ht]
\centering
    \resizebox{\columnwidth}{!}{%
    \includegraphics[scale=1]{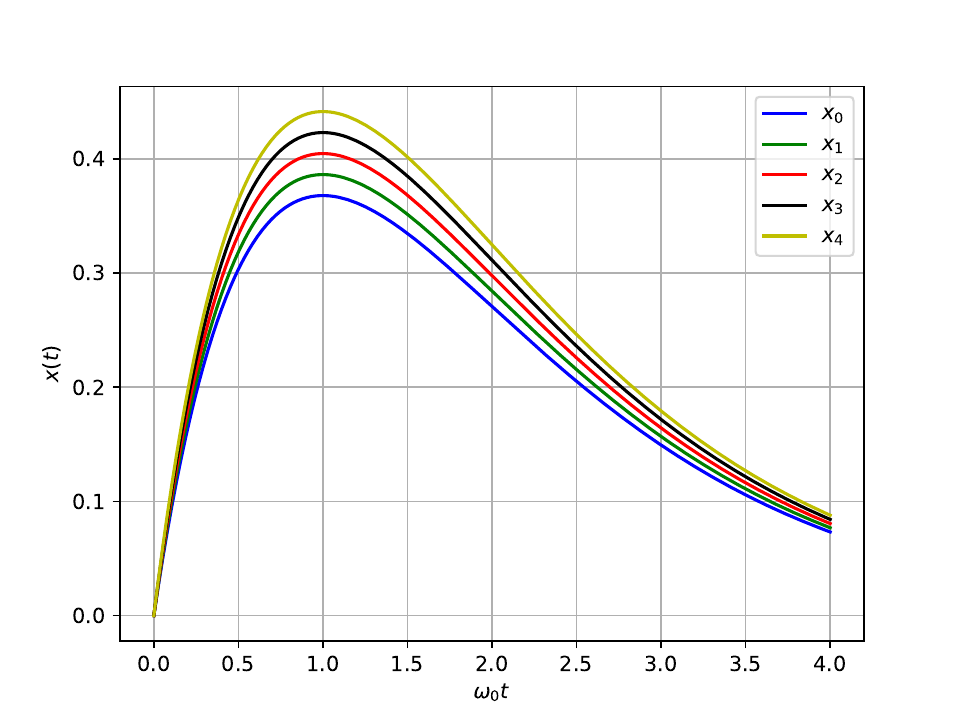}%
    }
\caption{Trajectories of a critically damped time-independent harmonic oscillator with slightly different initial velocities. See Table \ref{tab:Initial_conditions} for a list of these initial velocities. Here the displacement $x(t)$ is reported in arbitrary units.}
\label{fig:Damped_HO_x}
\end{figure}

To check this method, I numerically solved the equation of motion of the TDHO (Eq. \ref{eq:TDHO_EoM}) using the differential equation solver \textsc{odeint} packaged with \textsc{SciPy},\footnote{See \url{https://docs.scipy.org/doc/scipy/reference/generated/scipy.integrate.odeint.html} for a description of the \textsc{odeint} library.} and plotted the results in Fig. (\ref{fig:TDHO_x}). Each curve has an initial displacement of $x(t = 0) = 0$ and an initial velocity given in Table \ref{tab:Initial_conditions}, where
\begin{equation}
    v := \frac{dx}{d\left(\omega_0 t\right)}.
\end{equation}
From Fig. \ref{fig:TDHO_x}, we can see that trajectories having different initial velocities intersect for the first time around $\omega_0 t \approx 2.5$, in agreement with what Fig. \ref{fig:Ground_eig_var_est_python} predicts.\footnote{Like the undamped harmonic oscillator, the TDHO is a special case in that trajectories having arbitrarily different initial velocities will intersect at the kinetic focus; it is not necessary for the difference between the initial velocities of two trajectories to be small.} If one were to solve instead the equation of motion of a critically damped, time-independent harmonic oscillator, one would find the results shown in Fig. \ref{fig:Damped_HO_x}. In this case, trajectories with different initial velocities do not intersect, again in agreement with the prediction of Fig. \ref{fig:Ground_eig_var_est_python}.

These findings demonstrate that the dynamical stability of a non-conservative system can, under some circumstances, be qualitatively different from the dynamical stability of a conservative system, and that the Herglotz action is amenable to the same kind of stability analysis as the standard conservative action used in Ref. \citenum{Hussein_et_al_1980}.

\section{Conclusion}
\label{sec:conclusion}
The sign of the second variation of the action, for a conservative system, can be used as an indicator of that system's dynamical stability. As we have seen (and as shown in Refs. \citenum{Gray_Taylor_2007, Papastavridis_1983, Gray_Poisson_AJP_2011}, for example), the second variation changes sign when a conservative system passes through a kinetic focus, and it is the existence of these kinetic foci that indicate dynamical stability.

In this paper, I have generalized the analyses of Refs. \citenum{Gray_Taylor_2007} and \citenum{Hussein_et_al_1980} by showing that the second variation of the Herglotz action for a non-conservative system also changes sign when that system passes through a kinetic focus. This means that the second variation of the Herglotz action can be used to infer the dynamical stability properties of the class of non-conservative systems that can be described by the Herglotz action. Further, the eigenvalue analysis of the action for conservative systems described in Refs. \citenum{Hussein_et_al_1980} and \citenum{Levit_Smilansky_AnnPhys_1977} can also be applied to non-conservative systems. By doing this, I was able to locate the first kinetic focus of a TDHO, and I showed that the dynamical stability of this system, under a particular set of assumptions, is qualitatively different from the dynamical stability of a time-independent harmonic oscillator. It would be interesting to extend these results to more physically motivated applications of the TDHO, like those mentioned in Sec. \ref{subsec:SHO_g}.

Another way to extend these results would be to derive the semiclassical propagator, in the path integral formalism, for a quantum system that has dissipation and whose classical action is the Herglotz action. As I mentioned in the Introduction, the number of kinetic foci along a stationary (classical) path of a conservative quantum system determines the phase of the semiclassical propagator for that system. It would therefore be interesting to use the formalism developed here to investigate how dissipation affects the phase of the semiclassical propagator for a system whose dynamics can be derived from the Herglotz action. I leave this for future work.

\begin{acknowledgments}
I would like to thank Joel Meyers (SMU) for his detailed and insightful feedback at various stages of this project, and I wish to thank the anonymous referee for their helpful comments. Computational resources for this project were provided by SMU’s Center for Research Computing, and I used the following \textsc{Python} packages: \textsc{Matplotlib},\cite{Hunter_CSE_2007} \textsc{NumPy},\cite{Harris_et_al_2020} and \textsc{SciPy}.\cite{Virtanen_et_al_2020} I was financially supported, in part, by DOE grant DE-SC0010129. Data sharing is not applicable to this article as no new data were created or analyzed in this study. I have no conflicts of interest to disclose.
\end{acknowledgments}

\appendix

\section{Derivation of the action eigenvalue equation}
\label{app:A}
If we integrate the first term in parentheses on the RHS of Eq. (\ref{eq:d2A=}), we get
\begin{equation}
\label{eq:P_beforeintegration}
\begin{aligned}
    & \int_0^t\dot{y}_i P_{ij}\dot{y}_je^{-\int_0^{\tau_2} B d\tau_1}d\tau_2\\
    & = \int_0^t y_{i}\left[-\frac{d}{dt}\left(P_{ij}\dot{y}_j\right) + BP_{ij}\dot{y}_j\right]e^{-\int_0^{\tau_2} B d\tau_1}d\tau_2\\
    & = \int_0^t y_i e^{-\frac{1}{2}\int_0^{\tau_2} B d\tau_1}\left[-\frac{d}{dt}\left(P_{ij}\dot{y}_j\right) + BP_{ij}\dot{y}_j\right]e^{-\frac{1}{2}\int_0^{\tau_2} B d\tau_1}d\tau_2,\\
\end{aligned}
\end{equation}
and if we integrate the second term in parentheses on the RHS of Eq. (\ref{eq:d2A=}), we get
\begin{equation}
\label{eq:Q_beforeintegration}
\begin{aligned}
    & \int_0^t 2Q_{ij}y_i \dot{y}_j e^{-\int_0^{\tau_2}Bd\tau_1}d\tau_2\\
    & = \int_0^t y_i Q_{ij}\dot{y}_je^{-\int_0^{\tau_2}Bd\tau_1}d\tau_2 + \int_0^t y_j Q_{ji}\frac{dy_i}{dt}e^{-\int_0^{\tau_2}Bd\tau_1}d\tau_2\\
    & = \int_0^t y_i Q_{ij}\dot{y}_je^{-\int_0^{\tau_2}Bd\tau_1}d\tau_2 - \int_0^t y_j \frac{d}{dt}\left(Q_{ji}y_ie^{-\int_0^{\tau_2}Bd\tau_1}\right)d\tau_2\\
    & = \int_0^t y_i Q_{ij}\dot{y}_je^{-\int_0^{\tau_2}Bd\tau_1}d\tau_2\\
    & - \int_0^t \left(y_j\dot{Q}_{ji}y_i + y_jQ_{ji}\dot{y}_i - y_jBQ_{ji}y_i\right)e^{-\int_0^{\tau_2}Bd\tau_1}d\tau_2\\
    & = \int_0^t y_ie^{-\frac{1}{2}\int_0^{\tau_2}Bd\tau_1}\left(-\dot{Q}_{ji} + BQ_{ji}\right)y_je^{-\frac{1}{2}\int_0^{\tau_2}Bd\tau_1}d\tau_2.\\
\end{aligned}
\end{equation}
(recall the definitions of $P_{ij}$ and $Q_{ij}$ from Eqs. \ref{eq:P_def} and \ref{eq:Q_def}). On the other hand, if we insert Eqs. (\ref{eq:R_def}), (\ref{eq:Lambda_def}), and (\ref{eq:eta_def}) directly into Eq. (\ref{eq:d2A=Lambda}), we find
\begin{equation}
    \begin{aligned}
        \delta^2 \cA & = \int_0^t y_i e^{-\frac{1}{2}\int_0^{\tau_2}Bd\tau_1}\left[-\frac{d}{dt}\left(P_{ij}\frac{d}{dt} + Q_{ji}\right) + \left(Q_{ij}\frac{d}{dt} + R_{ij}\right)\right]y_j e^{-\frac{1}{2}\int_0^{\tau_2}Bd\tau_1}d\tau_2\\
        & = \int_0^t y_i e^{-\frac{1}{2}\int_0^{\tau_2}Bd\tau_1}\bigg\{\bigg[-\frac{d}{dt}\left(P_{ij}\dot{y}_{j}\right) + BP_{ij}\dot{y}_j\\
        & + \left(\frac{B}{2}\dot{P}_{ij} - \frac{B^2}{4}P_{ij} + \frac{1}{2}\frac{dB}{dt}P_{ij}\right)y_j\bigg] e^{-\frac{1}{2}\int_0^{\tau_2}Bd\tau_1} - \frac{dQ_{ji}}{dt}y_j e^{-\frac{1}{2}\int_0^{\tau_2}Bd\tau_1}\\
        & + \bigg(\frac{\partial^2 L}{\partial x_i \partial x_j} + 2\frac{\partial^2 L}{\partial x_i\partial S}\frac{\partial L}{\partial \dot{x}_j} + \frac{\partial^2 L}{\partial S^2}\frac{\partial L}{\partial \dot{x}_i}\frac{\partial L}{\partial \dot{x}_j}\\
        & + B Q_{ji} + \frac{B^2}{4}P_{ij} - \frac{1}{2}\frac{dB}{dt}P_{ij} - \frac{B}{2}\dot{P}_{ij}\bigg)y_j e^{-\frac{1}{2}\int_0^{\tau_2}Bd\tau_1}\bigg\}d\tau_2.\\
    \end{aligned}
\end{equation}
After canceling terms, this reduces to
\begin{equation}
\label{eq:d2A=ver3}
    \begin{aligned}
        \delta^2 \cA & = \int_0^t y_i e^{-\frac{1}{2}\int_0^{\tau_2}Bd\tau_1}\bigg\{\bigg[-\frac{d}{dt}\left(P_{ij}\dot{y}_{j}\right) + BP_{ij}\dot{y}_j\bigg]\\
        & + \bigg(\frac{\partial^2 L}{\partial x_i \partial x_j} + 2\frac{\partial^2 L}{\partial x_i\partial S}\frac{\partial L}{\partial \dot{x}_j} + \frac{\partial^2 L}{\partial S^2}\frac{\partial L}{\partial \dot{x}_i}\frac{\partial L}{\partial \dot{x}_j} + B Q_{ji} - \frac{dQ_{ji}}{dt}\bigg)y_j\bigg\}e^{-\frac{1}{2}\int_0^{\tau_2}Bd\tau_1}d\tau_2,\\
    \end{aligned}
\end{equation}
which is equal to the sum of Eqs. (\ref{eq:P_beforeintegration}), (\ref{eq:Q_beforeintegration}), and
\begin{equation}
\label{eq:no_eq}
    \int_0^t y_i e^{-\frac{1}{2}\int_0^{\tau_2}Bd\tau_1}\bigg(\frac{\partial^2 L}{\partial x_i \partial x_j} + 2\frac{\partial^2 L}{\partial x_i\partial S}\frac{\partial L}{\partial \dot{x}_j} + \frac{\partial^2 L}{\partial S^2}\frac{\partial L}{\partial \dot{x}_i}\frac{\partial L}{\partial \dot{x}_j}\bigg)y_je^{-\frac{1}{2}\int_0^{\tau_2}Bd\tau_1}d\tau_2.
\end{equation}
The RHS of Eq. (\ref{eq:d2A=}) is also equal to expression (\ref{eq:no_eq}) plus  Eqs. (\ref{eq:P_beforeintegration}) and (\ref{eq:Q_beforeintegration}), so it must equal Eq. (\ref{eq:d2A=ver3}). Therefore $\delta^2 \cA$ can be written in the form of Eq. (\ref{eq:d2A=Lambda}).

\section{Third variation of the Herglotz action}
\label{app:B}
$\tilde{A}''$ consists of various combinations of third derivative of $L$, of the form
\begin{equation}
    \tilde{A}'' \sim \frac{\partial^3 L}{\partial X_i \partial X_j \partial X_k}Y_i Y_j Y_k,
\end{equation}
where $X_i = x_i, \dot{x}_i$, or $S$, and $Y_i = y_i, \dot{y}_i$, or $f$. To lowest order
\begin{equation}
\label{eq:3derivL}
    \frac{\partial^3 L}{\partial X_i \partial X_j \partial X_k}\left(x_1\right) \approx \frac{\partial^3 L}{\partial X_i \partial X_j \partial X_k}\left(x_0\right) + \epsilon \frac{\partial^4 L}{\partial X_i \partial X_j \partial X_k \partial x_l}\left(x_0\right)y_l.
\end{equation}
In the Taylor expansion of the action, $\delta^3 S$ is associated with a factor of $\frac{\epsilon^3}{6}$. Therefore, to lowest order, the second term on the RHS of Eq. (\ref{eq:3derivL}) can be dropped in Eq. (\ref{eq:f''=}). In a similar fashion,
\begin{equation}
    B'' \sim \frac{\partial^3 L}{\partial X_i \partial X_j \partial S}Y_i Y_j,
\end{equation}
and
\begin{equation}
    \frac{\partial^3 L}{\partial X_i \partial X_j \partial S}\left(x_1\right) \approx \frac{\partial^3 L}{\partial X_i \partial X_j \partial S}\left(x_0\right) + \epsilon\frac{\partial^4 L}{\partial X_i \partial X_j \partial S \partial x_k}y_k.
\end{equation}
Also,
\begin{equation}
    \frac{\partial L}{\partial \dot{x}_i}\left(x_1\right) \approx \frac{\partial L}{\partial \dot{x}_i}\left(x_0\right) + \epsilon\frac{\partial L}{\partial \dot{x}_i \partial x_j}y_j
\end{equation}
As before, we can drop $O(\epsilon)$ terms from the product $B'' \frac{\partial L}{\partial \dot{x}_i}$. Finally, 
\begin{equation}
    B' \sim \frac{\partial^2 L}{\partial X_i \partial S}Y_i
\end{equation}
and
\begin{equation}
    \frac{\partial^2 L}{\partial X_i \partial S}\left(x_1\right) \approx \frac{\partial^2 L}{\partial X_i \partial S}\left(x_0\right) + \epsilon \frac{\partial^3 L}{\partial X_i \partial S \partial x_j}y_j,
\end{equation}
and we know that
\begin{equation}
\begin{aligned}
    f' & = e^{-\int B^{(0)} d\tau_3}\int_0^t \left[2\Omega(x_0)\right]e^{-\int_0^{\tau_2} B^{(0)} d\tau_1} d\tau_2 + O(\epsilon)\\
    & =: f'_{(0)} + O(\epsilon).\\
\end{aligned}
\end{equation}
Therefore
\begin{equation}
    f''(x_1) = e^{\int B d\tau_3}\left[\int_0^t \left(\tilde{A}'' + B''\frac{\partial L}{\partial \dot{x}_i}y_i + 2B'f'\right)e^{-\int_0^{\tau_2}Bd\tau_1}d\tau_2\right] + O(\epsilon)
\end{equation}
and
\begin{equation}
    f''(x_0) = e^{\int B d\tau_3}\left[\int_0^t \left(\tilde{A}'' + B''\frac{\partial L}{\partial \dot{x}_i}y_i + 2B'f'\right)e^{-\int_0^{\tau_2}Bd\tau_1}d\tau_2\right],
\end{equation}
(with $B$, $\tilde{A}''$, $B'$, $B''$, $f'$, and $\frac{\partial L}{\partial \dot{x}_i}$ in both equations evaluated on $x_0$).

\bibliography{aiptemplate}

\end{document}